\documentclass[
pra, 
aps,
twocolumn,
superscriptaddress,
longbibliography,
floatfix,
notitlepage]{revtex4-2}

\usepackage[utf8]{inputenc}
\usepackage[english]{babel}
\usepackage[T1]{fontenc}
\usepackage{lmodern}

\usepackage{amsmath,amsfonts,amssymb,amsthm,bm,times,dcolumn}

\usepackage{microtype}
\usepackage{braket}
\usepackage{gensymb} % for example \degree symbol can be used!
\usepackage{physics}

\usepackage[colorlinks={true}, citecolor={blue}, filecolor={blue}, 
linkcolor={blue}, urlcolor={blue}]{hyperref}
\usepackage{graphicx,color}
\usepackage[caption=false]{subfig}

\usepackage{soul}
\usepackage[normalem]{ulem}

\usepackage{longtable}
\usepackage{supertabular}
\begin{document}
	
\preprint{APS/123-QED}

%\title{Deterministic preparation of Dicke states via collisions}
\title{Intelligent Control of Collisional Architectures for Deterministic Multipartite State Engineering}

\author{Duc-Kha Vu}
\affiliation{KT One, 4302 Shire Court, Tampa, FL, 33613, United States}
\affiliation{Tokyo International University, 4-42-31 Higashi-Ikebukuro, Toshima-ku, Tokyo, 170-0013, Japan}

\author{Minh Tam Nguyen}
\affiliation{KT One, 4302 Shire Court, Tampa, FL, 33613, United States}
\affiliation{University of South Florida, 4202 E Fowler Ave, Tampa, FL 33620, United States}
	
\author{\"{O}zg\"{u}r E. M\"{u}stecapl{\i}o\u{g}lu}
\affiliation{Department of Physics, Ko\c{c} University, Sar{\i}yer, \.Istanbul, 34450, Türkiye}
\affiliation{TÜB\.ITAK Research Institute for Fundamental Sciences (TBAE), 41470 Gebze, Türkiye}

\author{Fatih Ozaydin}
\email{mansursah@gmail.com}
\affiliation{Tokyo International University, 4-42-31 Higashi-Ikebukuro, Toshima-ku, Tokyo, 170-0013, Japan}
\affiliation{Nanoelectronics Research Center, Kosuyolu Mah., Lambaci Sok., Kosuyolu Sit., No:9E/3  Kadikoy, Istanbul, 34718, T\"urkiye}
	
\date{\today}

\begin{abstract}
	Designing scalable, noise-tolerant control protocols for multipartite entanglement is a central challenge for quantum technologies, and it naturally calls for \emph{algorithmic} synthesis of interaction parameters rather than handcrafted gate sequences. Here we introduce an intelligent, constraint-aware control framework for deterministic generation of symmetric Dicke states $|D_n^{(m)}\rangle$ in repeated-interaction (collision-model) architectures. The protocol employs excitation-preserving partial-SWAP collisions between two disjoint qubit registers, mediated by $m$ ancillary ``shuttle'' qubits, and poses Dicke-state preparation as a \emph{closed-loop design} problem: given the target $(n,m)$, automatically infer collision strengths that maximize fidelity under practical constraints.	
	Concretely, we formulate a two-parameter, bound-constrained optimization over intra-register and shuttle--register collision angles and solve it using a multi-start strategy with L-BFGS-B, yielding a reproducible controller prescription (optimized $\gamma_{\mathrm{in}}$, $\gamma_{\mathrm{sh}}$, and minimal-round convergence points) for each target. This removes the need for projective measurements and extends collisional entanglement generation beyond the single-excitation (W-state) sector to arbitrary $m$.	
	Crucially, we optimize \emph{within} imperfect collisional dynamics where errors act throughout the sequence, including stochastic interaction dropouts (missing collisions) and standard decoherence channels. Strikingly, across wide error ranges the optimized controller preserves high preparation fidelity; imperfections manifest primarily as a modest increase in the required number of collision rounds. This behavior reflects a tunable competition in which noise suppresses correlations while properly chosen collisions continuously replenish them, allowing the control algorithm to trade time for fidelity. Our results establish an automated and physically transparent route to robust, deterministic Dicke-state engineering, and provide a concrete blueprint for integrating collision-model state synthesis with emerging quantum platforms.
\end{abstract}

\maketitle

%%%%%%%%%%%%%%%%%%%%%%%%%%%%%%%%%%%%%%%%%%%%%%%%%%%%%%%%%%%%%%%%%%%%%
\section{Introduction}\label{sec:Introduction}
%%%%%%%%%%%%%%%%%%%%%%%%%%%%%%%%%%%%%%%%%%%%%%%%%%%%%%%%%%%%%%%%%%%%%

Multipartite entanglement is a central resource for quantum technologies, yet its reliable preparation in scalable architectures remains a persistent challenge. Among the most prominent families of multipartite states are the Dicke states, i.e., permutation-symmetric states with a fixed number of excitations distributed over $N$ qubits. Originally introduced in the context of cooperative light--matter emission, Dicke states provide the natural language for describing collective radiative phenomena such as superradiance and subradiance, where indistinguishable emitters exhibit enhanced or suppressed spontaneous emission due to symmetry and quantum interference~\cite{dicke1954coherence,rehler1971superradiance,gross1982superradiance}. In modern settings, Dicke physics also underpins a broad range of effects in collective spin models and waveguide QED, and continues to motivate state engineering in platforms ranging from atomic ensembles and trapped ions to superconducting circuits~\cite{amsuss2011cavity,sheremet2023waveguide}.

Beyond their foundational role in superradiant dynamics, Dicke states are widely recognized as versatile resources for quantum information processing. They support metrological advantages in distributed sensing and phase estimation (particularly in symmetric subspaces)~\cite{hotter2024combining,saleem2024achieving}, enable robust multipartite entanglement detection via collective observables~\cite{lohof2023signatures, kiesel2007experimental, lucke2014detecting}, and provide structured entanglement useful for quantum networking protocols and multipartite communication tasks~\cite{chiuri2012experimental,cheng2020realizing}. Very recently, Dicke states have also been investigated from a topological perspective~\cite{bhattacharyya2025entanglement}.
Special cases include $W$ states (single-excitation Dicke states), which are known for their characteristic robustness of pairwise entanglement under particle loss~\cite{neven2018entanglement} and decoherence~\cite{sen2003multiqubit}, and multi-excitation Dicke states which offer richer correlation structure and tunable entanglement depth~\cite{chen2016entanglement}. These advantages have motivated a substantial literature on Dicke-state preparation.

While the preparation of GHZ and cluster states is relatively straightforward~\cite{nielsen2010quantum}, the generation of Dicke states, even with single-excitation, i.e., $W$ states, has historically required substantial effort. A variety of approaches have been proposed, including probabilistic fusion of smaller $W$ states to obtain larger ones~\cite{bugu2013enhancing, yesilyurt2013optical, ozaydin2014fusing, bugu2020, li2016generating, zang2015generating}, as well as deterministic expansion protocols that enlarge a seed $W$ state by coupling it to separable ancillary qubits~\cite{ozaydin2021deterministic,yesilyurt2016deterministic,zang2016deterministic}.

Proposed approaches for preparing Dicke states with an arbitrary number of excitations include global control~\cite{stojanovic2023dicke}, phase-estimation-based methods~\cite{wang2021preparing}, adiabatic and counterdiabatic protocols~\cite{carrasco2024dicke,opatrny2016counterdiabatic}, engineered dissipation~\cite{zhu2025dissipation}, and geometric schemes~\cite{yu2024efficient}. Circuit-based constructions have also been developed to reduce gate complexity~\cite{bartschi2019deterministic,dutta2024following,mukherjee2020preparing,aktar2022divide}, including scenarios with limited qubit access~\cite{thapa2025expanding}.

Collision (or repeated-interaction) models~\cite{ciccarello2022quantum, carruthers1983quantum, campbell2021collision} have attracted increasing attention in recent years. It has been shown that they can efficiently simulate arbitrary Markovian quantum dynamics~\cite{cattaneo2021collision} and enable synchronization beyond the Markovian limit in dissipative environments~\cite{karpat2021synchronization}. Collision models have also provided valuable insights in quantum thermodynamics~\cite{de2020quantum, maity2025violation, manatuly2019collectively, pedram2022environment} and quantum Darwinism~\cite{campbell2019collisional}. Beyond these applications, they offer a physically transparent paradigm for entanglement generation in open quantum systems. In their simplest form, a system interacts sequentially with ancillary units (``environmental'' qubits) through a fixed two-body unitary, enabling a controlled build-up of correlations via repeated interactions~\cite{ziman2010open}.

A particularly appealing variant of collision models introduces a mobile ancillary qubit (a ``shuttle'') that mediates indirect interactions among otherwise noninteracting register qubits through energy-preserving partial-SWAP collisions. This mechanism was shown to generate genuine multipartite entanglement across multiple registers, and in the four-qubit case to prepare a $W$ state with enhanced performance when supplemented by projective measurements on the shuttle~\cite{ccakmak2019robust}. While measurement-assisted postselection can boost entanglement, it also renders the protocol intrinsically probabilistic and complicates integration into deterministic quantum-information workflows.

In this work, we develop a deterministic and noise-resilient protocol for preparing Dicke states via collision dynamics. Our key idea is to treat the collision strengths (equivalently, interaction times) as design parameters and to optimize them algorithmically so that the register converges to a targeted Dicke state without any projective measurements on the shuttle. This generalizes measurement-assisted $W$-state generation to the broader class of multi-excitation Dicke states and establishes a constructive route toward deterministic preparation within the same energy-preserving collision framework. Importantly, we assess a more realistic noise scenario in which decoherence acts throughout the collision sequence, rather than being applied only after the idealized unitary dynamics. We find that optimized collision parameters can substantially mitigate the net degradation of multipartite entanglement by continuously replenishing correlations during the noisy evolution, thereby enabling robust Dicke-state generation in parameter regimes where non-optimized protocols fail.

{\color{black}
The remainder of the paper is organized as follows. Section~\ref{sec:Preliminaries} summarizes background and notation for multi-excitation Dicke states and clarifies why the preparation task becomes more demanding beyond the single-excitation (W-state) sector. In Section~\ref{sec:Methods}, we present the collision-model architecture based on excitation-preserving partial-SWAP interactions, introduce the fidelity-based objective, and describe our bound-constrained optimization workflow (multi-start initialization combined with L-BFGS-B) used to determine the optimal collision strengths. Section~\ref{sec:Results} reports the deterministic state-preparation performance for Dicke targets up to ten qubits and quantifies robustness when imperfections act throughout the collision sequence, including missing-collision events and standard decoherence channels. In Section~\ref{sec:Discussions}, we discuss practical aspects of the protocol, including the impact of grid resolution and computational cost, the observed (non-monotonic) scaling of collision rounds under imperfections, and possible physical implementation routes in both analog exchange-based platforms and digital quantum-simulation settings. Finally, Section~\ref{sec:Conclusions} concludes with a summary and an outlook.
}

\section{Preliminaries}\label{sec:Preliminaries} 
A three-qubit Dicke state in the single- or double-excitation sector is, up to local operations and classical communication (LOCC), equivalent to a $W$ state. In contrast, already at four qubits the two-excitation Dicke state,
\begin{eqnarray}\label{eq:Dicke42} 
	|D_4^{(2)}\rangle & = &\frac{1}{\sqrt{6}} ( |1100\rangle + |1010\rangle + |1001\rangle \\ \nonumber 
	& & + |0110\rangle + |0101\rangle + |0011\rangle ) 
\end{eqnarray}
supports a qualitatively richer pattern of multipartite correlations. We denote an $n$-qubit Dicke state with $m$ excitations by $|D_n^{(m)}\rangle$, so that $|D_n^{(1)}\rangle\equiv |W_n\rangle$. Unlike $W$ states, whose entanglement structure is constrained by the presence of a single delocalized excitation, multi-excitation Dicke states distribute quantum correlations over a much larger set of computational-basis configurations. This enlarged excitation manifold can support higher entanglement depth and more intricate multipartite correlation patterns. At the same time, the increased complexity of the target state typically translates into a more demanding state-engineering task: if preparing $|W_n\rangle$ has already been challenging, generating $|D_n^{(m)}\rangle$ for arbitrary $m$ is, in general, even more challenging.

\section{Methods}\label{sec:Methods} 

\subsection{Collision Model}\label{sec:CollisionModel}
\begin{figure}[t!]
	\includegraphics[width=1\columnwidth]{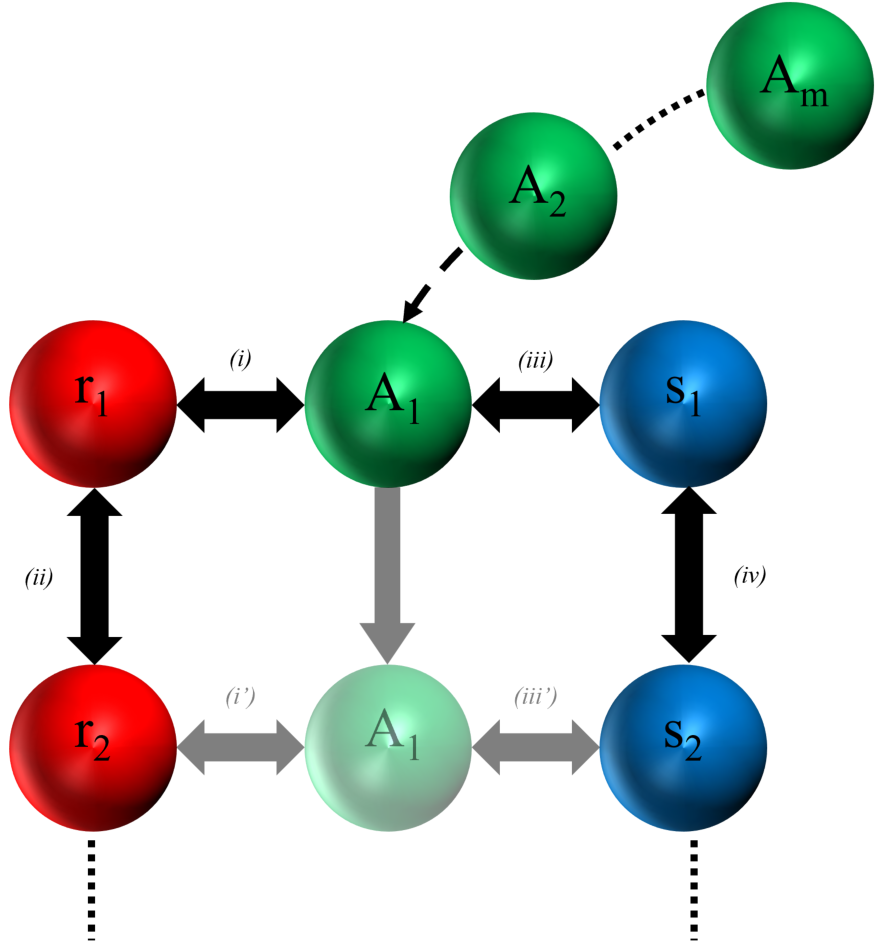}
	\caption{Collision scheme for preparing $|D_n^{(m)}\rangle$, i.e., an $n$-qubit Dicke state with $m$ excitations. In the case of $n=6$ and $m=2$, the procedure of the shuttle qubit $A_1$ consisting of discretized  interaction steps (i, ii, iii, iv, i', ii, iii', iv') is repeated by the $A_2$ qubit.
		\label{fig:fig_model}}
\end{figure}

We consider two disjoint registers of qubits, $r$ and $s$, composed of qubits $r_i$ and $s_i$ with $i \in {1,2,\ldots,l}$, and a set of ``shuttle'' qubits $A_j$ with $j \in {1,2,\ldots,m}$, where $m$ is the targeted number of excitations, and $n = 2l+m$. We illustrate the collision model in Fig.~\ref{fig:fig_model} for preparing a an arbitrary Dicke state, with the procedure for a $|D_6^{(2)}\rangle$ state as an example: each register contains $l=2$ qubits and the protocol employs $m=2$ shuttle qubits. The register qubits interact with their nearest neighbors, and the shuttle qubits interact with qubits in the disjoint registers following the procedure.

With $\gamma$ representing the interaction duration, or equivalently the interaction strength, we model each collision as a two-qubit partial SWAP operator
\begin{equation}\label{eq:partial_swap}
	U(\gamma) =	\cos(\gamma) \mathbb{I} + i \sin(\gamma) \text{SWAP},
\end{equation}
\noindent
which reduces to the two-qubit identity operator $\mathbb{I}$ and to the two-qubit SWAP operator
\begin{equation}
	\mathrm{SWAP} =
	\begin{pmatrix}
		1 & 0 & 0 & 0 \\
		0 & 0 & 1 & 0 \\
		0 & 1 & 0 & 0 \\
		0 & 0 & 0 & 1
	\end{pmatrix},
	\label{eq:SWAP}
\end{equation}
for $\gamma=0$ and $\gamma=\pi / 2$, respectively. 

The interaction between the shuttle and register qubits is modeled as an energy-preserving exchange. 
While the full Heisenberg exchange interaction is given by $J\tau(\vec{\sigma}_A \cdot \vec{\sigma}_i)$~\cite{ccakmak2019robust}, 
the deterministic preparation of Dicke states relies specifically on the redistribution of excitations within the $m$-excitation subspace. Following the effective model described in Ref.~\cite{ccakmak2019robust}, we omit the longitudinal $\sigma_z$ components which only contribute to a phase shift, and focus on the transverse ladder operators that drive the necessary excitation exchange. Consequently, the partial SWAP operation is physically realized via the interaction:
\begin{equation}
	\text{SWAP} \approx \exp\left[i \gamma (\sigma_+^A \sigma_-^a + \sigma_-^A \sigma_+^a)\right], \label{eq:RevisedSwap}
\end{equation}
where $\sigma_\pm = \frac{1}{2}(\sigma_x \pm i\sigma_y)$ are the qubit raising and lowering operators, and $\sigma^{a}$ is with $a = r, s$. 
This formulation ensures that the total excitation number is strictly conserved throughout the collision sequence, 
facilitating the convergence of the register to the target $|D_{n}^{(m)}\rangle$ state.

In Ref.~\cite{ccakmak2019robust}, the shuttle--register interaction strength was fixed and for the intra-register interaction strength, only two values were considered: $\gamma = 0$, and $\gamma = 0.95 \pi /2$ corresponding to vanishing and strong interactions, respectively. A consequence of this is that the protocol was formulated with a single shuttle qubit, which restricts the accessible target manifold, addressing only the $W$-state generation. A second consequence is the reliance on a projective measurement on the shuttle: besides tracing the shuttle out of the final multipartite state (thereby reducing the size of the prepared state), this measurement renders the protocol probabilistic, with a success probability that decreases as $n$ increases.

In this work, we treat both interaction strengths as free parameters, denoted by $\gamma_{\mathrm{sh}}$ and $\gamma_{\mathrm{in}}$ for shuttle--register and intra--register collisions, respectively. This extended control enables \textit{i)} the use of not only a single shuttle but $m$ shuttle qubits to realize the desired excitation number, \textit{ii)} the elimination of projective measurements on the shuttle qubit(s), thereby yielding a deterministic protocol that prepares the target Dicke state of $n$ qubits and $m$ excitations with very high fidelity, and \textit{iii)} enhanced robustness against decoherence in a more realistic setting than that considered in Ref.~\cite{ccakmak2019robust}: while decoherence continuously degrades entanglement, appropriately optimized collisions can continuously re-entangle the qubits and mitigate the net loss, allowing high-fidelity preparation even in the presence of realistic noise.

	Preparing a Dicke state under repeated collisions is analogous to a controlled \emph{mixing-and-symmetrization} process within a fixed-excitation manifold. Each partial-SWAP collision coherently redistributes excitations between the shuttle and the register qubits, so that population and phase weight are repeatedly transferred among computational-basis configurations with the same total excitation number. Over many rounds, this repeated redistribution can be viewed as an entanglement-pumping mechanism that explores the relevant subspace, while the choice of collision strengths determines the interference pattern among the contributing amplitudes. Our optimization procedure exploits this structure by tuning $(\gamma_{\mathrm{in}},\gamma_{\mathrm{sh}})$ so that the stroboscopic evolution amplifies the symmetric Dicke component and suppresses undesired components, effectively steering the register toward $|D_n^{(m)}\rangle$.

\subsection{Optimizing Interaction Strengths}\label{sec:EstimatingInteractionStrengths}

Allowing both collision strengths to vary introduces a nontrivial optimization problem: identifying near-optimal parameter values becomes increasingly challenging for large $n$ and arbitrary excitation number $m$. To address this challenge, we define a loss function based on the fidelity between the generated state and the ideal target state. Since both interaction strengths correspond to real angles in Eq.~\ref{eq:partial_swap}, we impose the boundary constraints $\gamma_{\mathrm{in}} \in [0,\pi]$, and $\gamma_{\mathrm{sh}} \in [0.01,\pi]$ (because $\gamma_{\mathrm{sh}}=0$ implies no shuttle collisions). For a finite number of collisions, we then solve the resulting bound-constrained optimization problem using the limited-memory BFGS algorithm with bound constraints (L-BFGS-B)~\cite{byrd1995limited}.

We initialize our setup with $m$ shuttle qubits ($A$) and $n-m=2l$ register qubits ($rs$) as 
\begin{align}
	|\psi_0\rangle
	= |1\rangle^{\otimes m}_A \otimes |0\rangle^{\otimes (n-m)}_{rs},
\end{align}
and consider the quantum operation for one round of collisions illustrated in Fig.~\ref{fig:fig_model} as
\begin{align}
	\mathcal{U}_{\mathrm{round}}
	= U_{\mathrm{iv}}U_{\mathrm{iii'}}
	U_{\mathrm{ii}}U_{\mathrm{i'}}
	U_{\mathrm{iv}}U_{\mathrm{iii}}
	U_{\mathrm{ii}}U_{\mathrm{i}},
\end{align}
or in a concise way
\begin{align}
	\mathcal{U}_{\mathrm{round}}
	= U_{\mathrm{intra}}(\gamma_{\mathrm{in}}) \ 
	U_{\mathrm{shuttle}}(\gamma_{\mathrm{sh}}).
\end{align}
After $r$ rounds, the system evolves as
\begin{align}
	|\psi_r\rangle
	=
	\bigl(\mathcal{U}_{\mathrm{round}})^r
	|\psi_0\rangle.
\end{align}
In order to estimate the two control angles $(\gamma_{\mathrm{in}}$ and $\gamma_{\mathrm{sh}})$ by minimizing the fidelity-based loss between the generated state and the ideal Dicke state with the desired excitations, let $R$ be the maximum number of rounds to be executed. The non-zero terms of the target amplitude vector (of the $n$-qubit pure Dicke state) on the $m$-excitation subspace are equal to

\begin{align}
	\mathbf{a}_\star \;=\; \frac{1}{\sqrt{\binom{n}{m}}} .
\end{align}

\begin{figure}[t!]
	\includegraphics[width=1\columnwidth]{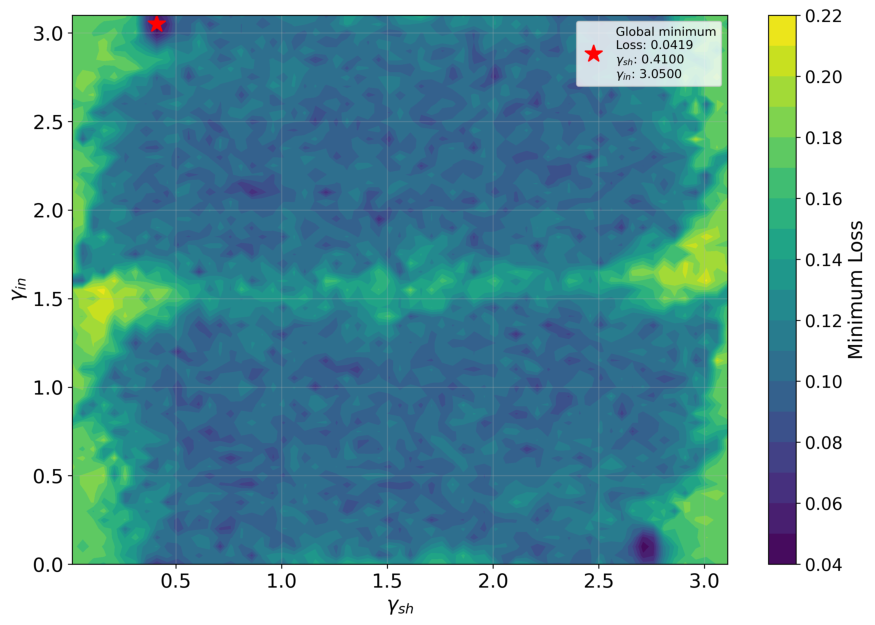}
	\caption{Minimal loss as a function of shuttle-register qubit interaction strength ($\gamma_{\mathrm{sh}}$) and intra-register qubit interaction strength ($\gamma_{\mathrm{in}}$). Red star denotes the optimal values yielding the minimum loss function based on the fidelity between the generated and ideal Dicke state, estimated by the L-BFGS-B simulation. 
		\label{fig:fig_sim_2D}}
\end{figure}

For identifying the optimal interaction strengths, we simulate the collision dynamics up to $R$ rounds and extract the amplitude vector $\mathbf{a}_r(\gamma_{\mathrm{in}} , \gamma_{\mathrm{sh}})$ on this subspace. The fidelity turns out to be
% a_r should be in abs
\begin{align}
	F_r(\gamma_{\mathrm{in}} , \gamma_{\mathrm{sh}}) \;=\; \bigl|\langle \mathbf{a}_\star| \mathbf{a}_r(\gamma_{\mathrm{in}} , \gamma_{\mathrm{sh}})\rangle \bigr|^2,
\end{align}
and the objective is to yield is the minimum loss across rounds,
\begin{align}
	\mathcal{L}(\gamma_{\mathrm{in}} , \gamma_{\mathrm{sh}}) \;=\; \min_{0 < r \le R}\, \bigl(1 - F_r(\gamma_{\mathrm{in}} , \gamma_{\mathrm{sh}})\bigr).
\end{align}

To reduce the sensitivity to local minima, we employ a dense multi-start strategy: initial guesses are taken from a uniform grid with spacing $0.2$ radians over $\gamma_{\mathrm{in}}\in[0,\pi]$ and $\gamma_{\mathrm{sh}}\in[0.01,\pi]$.
Using the L-BFGS-B with finite-difference gradients, we solve the bound-constrained problem
\begin{align}
	\min_{\gamma_{\mathrm{in}} , \gamma_{\mathrm{sh}} \in \mathcal{B}} \; \mathcal{L}(\gamma_{\mathrm{in}} , \gamma_{\mathrm{sh}}),
	\qquad
	\mathcal{B} \;=\; [0,\pi]\times[0.01,\pi].
\end{align}

For each point on the grid, we run the simulation for a total of $R$ rounds and then invoke L-BFGS-B from that grid point to iteratively search for a local minimizer of the loss. All runs are executed in parallel across CPU cores. Since L-BFGS-B is launched from each grid point, refining the grid by decreasing the spacing increases the number of initializations and, consequently, the computational cost approximately quadratically.

	In the simulations, we set $R = 200 \times m$ and employed the L-BFGS-B optimizer with a maximum number of iterations (\texttt{maxiter}) set to 100 
	and a function-tolerance stopping criterion (\texttt{ftol}), i.e., the minimum required decrease in the objective function between iterations, set to $10^{-6}$
	together with the strict box constraints specified above. 	
Accordingly, for each initialization point, L-BFGS-B performs at most 100 iterations, or terminates earlier if the improvement in the loss falls below $10^{-6}$. Throughout the procedure, the incumbent best $(\gamma^\star, r^\star, \mathcal{L}^\star)$ is updated whenever a lower-loss solution is found.

The estimated collision strengths only offer the best amplitude distribution across the basis states of the desired Dicke state without taking into account the relative phases among them. To address this, we perform a secondary optimization for each candidate pair of collision strengths to align the relative phases. This is achieved by applying local single-qubit $R_z$ rotations on each qubit, parameterized by rotation angles, which are subsequently optimized using the L-BFGS-B algorithm.

In Fig.~\ref{fig:fig_sim_2D}, we illustrate the optimization procedure: using the L-BFGS-B algorithm, we estimate the optimal values of the collision strengths yield the minimum loss based on the fidelity. Note that this figure is illustrating the simulation procedure only on the grid points, while the actual simulation runs for 100 points around each grid point.

\subsection{Robustness Analysis}\label{sec:RobustnessAnalysis}
To account for experimentally relevant conditions, we first focus on the interaction failures (missed collisions).

Beyond intrinsic decoherence, long collision sequences are also sensitive to operational errors in activating the intended two-qubit interaction. In repeated-interaction architectures where the coupling is mediated by a mobile ancilla or by a tunable coupler switched on for short time windows, a scheduled shuttle--register collision may effectively be skipped due to timing jitter, imperfect activation of the coupling pulse, or ancilla-transfer failures. To capture this, we model each intended collision as a stochastic \emph{gate-dropout} event: with probability $p_{\mathrm{miss}}$ the interaction is not enacted (identity operation), and with probability $1-p_{\mathrm{miss}}$ the ideal partial-SWAP collision $U(\gamma)$ is applied. Equivalently, the state update for an intended collision step is described by the completely positive trace-preserving map

\begin{equation}
	\mathcal{E}_{\mathrm{miss}}(\rho)=(1-p_{\mathrm{miss}})\,U(\gamma)\rho U^\dagger(\gamma)+p_{\mathrm{miss}}\,\rho.
\end{equation}

In an ensemble-averaged description, such random skip events introduce stochastic disorder in the interaction history and can manifest as dephasing-like phase slips in the accumulated amplitudes; any intrinsic single-qubit phase noise is captured separately by the dephasing channel below, which will be also examined explicitly.

The second imperfection class is the basic decoherence channels, i.e., dephasing, depolarizing, and amplitude damping with the associated decoherence probability $q$. Kraus operators of the dephasing channel are given as~\cite{nielsen2010quantum}

\begin{equation} 
	K^{\text{Deph}}_1 = \sqrt{1-q} 
	\begin{pmatrix} 
		1 & 0 \\ 
		0 & 1 
	\end{pmatrix}, \quad 
	K^{\text{Deph}}_2 = \sqrt{q} 
	\begin{pmatrix} 
		1 & 0 \\ 
		0 & -1 
	\end{pmatrix}, \label{eq:DephasingKraus} 
\end{equation}

\noindent implying that, with probability $1-q$, the qubit remains unchanged, whereas with probability $q$ its phase is flipped.

Depolarizing noise is a convenient effective model for incoherent control errors that randomly apply Pauli errors, and it is often used as a coarse-grained description of gate imperfections when multiple error sources act simultaneously. Physically, it can arise from pulse-amplitude and phase miscalibration, leakage and subsequent return to the computational subspace, crosstalk, or from fast stochastic fluctuations that, when averaged over, act isotropically on the Bloch sphere. While not intended as a microscopic model of a specific device, depolarization provides a conservative benchmark for the protocol under generic, unstructured noise, which is described by Kraus operators as

\begin{eqnarray}
	K^{\text{Depol}}_1 = \sqrt{1-3q/4 \ } \ \mathbb{I}, \quad K^{\text{Depol}}_2 = \sqrt{q \ } \ X/2, \\ K^{\text{Depol}}_3 = \sqrt{q \ } \ Y/2, \quad K^{\text{Depol}}_4 = \sqrt{q \ } \ Z/2,  \label{eq:DepolarizationKraus} 
\end{eqnarray}

\noindent where $X$, $Y$, and $Z$ are the identity, and the Pauli spin operators, respectively.

Amplitude damping captures energy relaxation processes in which an excited qubit irreversibly decays to its ground state by emitting energy into uncontrolled environmental modes. In solid-state and circuit-QED platforms this is typically associated with finite $T_1$ relaxation due to dielectric loss, Purcell-enhanced decay through the readout resonator, or coupling to spurious two-level systems, while in atomic and photonic settings it can arise from spontaneous emission or loss from the relevant two-level manifold. In the present protocol, such relaxation events directly reduce the excitation number and therefore compete with the preparation of $|D_n^{(m)}\rangle$, making amplitude damping a stringent and experimentally relevant robustness test. Its Kraus operators are given as

\begin{equation} 
	K^{\text{Damp}}_1 = 
	\begin{pmatrix} 
		1 & 0 \\ 
		0 & \sqrt{1-q}  
	\end{pmatrix}, \quad 
	K^{\text{Damp}}_2 = 
	\begin{pmatrix} 
		0 & \sqrt{q}  \\ 
		0 & 0
	\end{pmatrix}. \label{eq:AmplitudeDampingKraus} 
\end{equation}

For each imperfection scenario considered separately, we rerun the optimization procedure and identify, within the prescribed parameter bounds and grid spacing, the set of values that yields the highest final-state fidelity.\\

\begin{figure}[t!]
	\includegraphics[width=1\columnwidth]{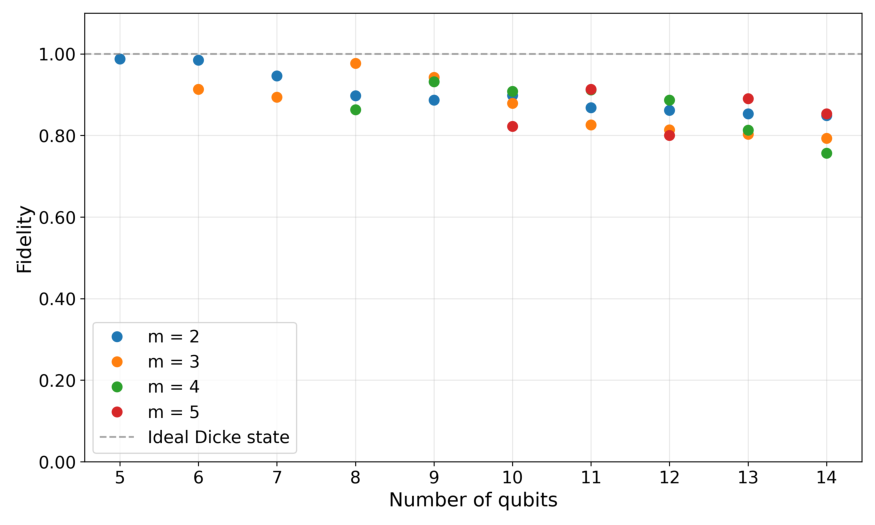}
	\caption{Fidelity between the generated and the ideal Dicke states for sizes up to 14 qubits, across the accessible excitation sectors. Here, the simulation setup is based on a uniform grid with spacing $0.2$ radians over $\gamma_{\mathrm{in}}\in[0,\pi]$ and $\gamma_{\mathrm{sh}}\in[0.01,\pi]$.
		\label{fig:fig_fidelity_10_qubits}}
\end{figure}

\vfill
\section{Results}\label{sec:Results} 

First, we benchmark the protocol in the ideal setting by running the full optimization workflow with a grid spacing of $0.2$ radians, targeting Dicke states up to $n=14$ qubits across various excitation numbers. For each pair $(n,m)$, the algorithm identifies collision strengths that maximize the final-state fidelity to the ideal target $|D_n^{(m)}\rangle$. The resulting fidelities are summarized in Fig.~\ref{fig:fig_fidelity_10_qubits}, which demonstrates that high-fidelity preparation is achievable throughout the considered state family. In Section~\ref{sec:Discussions} we further examine the role of grid resolution and show how narrower spacing impacts both the achievable fidelities and the associated computational cost.

We then turn to robustness and assess two experimentally motivated imperfections. The first is \emph{shuttle failure} (missing collisions), modeled by assigning a failure probability to each scheduled shuttle--register interaction: in a given round, the shuttle may fail to collide with the designated register qubit, effectively skipping that unitary step. We evaluate the protocol over a range of failure probabilities and, for each value, rerun the optimization to determine the best-performing collision strengths under the corresponding stochastic dynamics.

Since the optimization is explicitly performed \emph{within} the imperfect model, it naturally adapts the effective entangling schedule to the reduced interaction rate. As the failure probability increases, the dominant effect is a slowdown of entanglement accumulation rather than a qualitative breakdown of the mechanism: the target correlations can still be built up, albeit typically requiring a larger number of rounds to compensate for the missed interactions.
In addition to the aggregate fidelities reported in Fig.~\ref{fig:fig_fidelity_10_qubits}, we provide the underlying optimal settings for each $|D_n^{(m)}\rangle$ target. namely $\gamma_{\mathrm{sh}}$, $\gamma_{\mathrm{in}}$, the round count at which the best fidelity occurs, and the associated local $R_z$ angles used for phase alignment in Table~I in the Appendix.
\newpage

\onecolumngrid
\onecolumngrid
%\begin{widetext}
	\begin{figure}[h!]
		\includegraphics[width=0.99\columnwidth]{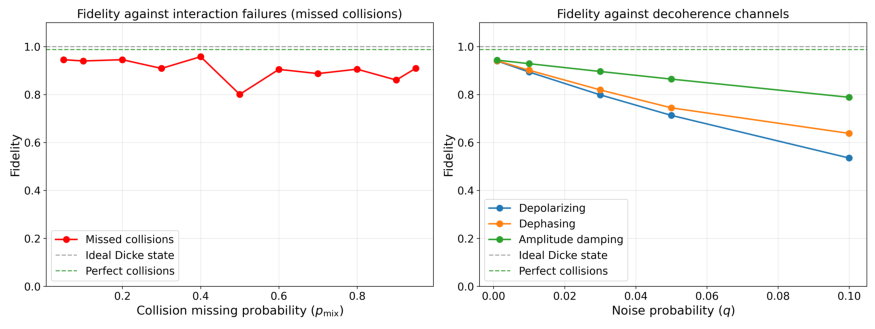}
		\caption{Robustness of the optimized collision protocol under experimentally motivated imperfections acting during the repeated-interaction dynamics for preparing $|D_5^{(2)}\rangle$ as an example. (a) Missing shuttle–register collisions: each scheduled shuttle–register interaction is skipped with a prescribed probability. (b) Decoherence channels (depolarizing, dephasing, amplitude damping): each independent noise channel acts throughout the collision process. For each imperfection strength, we rerun the bound-constrained optimization over $(\gamma_{\mathrm{in}},\gamma_{\mathrm{sh}})$ and report the maximum achieved fidelity with the target Dicke state (markers), alongside the corresponding noiseless baselines (dashed) and the ideal value $F=1$ (dotted).\label{fig:fig_robustness_analysis}}
	\end{figure}
%\end{widetext}
\twocolumngrid
\twocolumngrid

Next, we investigate decoherence acting during the collision process. In contrast to a simplified setting where decoherence is applied only after an idealized unitary evolution, here the decoherence continuously competes with entanglement generation throughout the repeated interactions. The resulting dynamics exhibits a clear interplay between two opposing tendencies: decoherence suppresses coherences and reduces multipartite entanglement, while the ongoing collisions repeatedly reintroduce coherences by redistributing excitations and generating fresh correlations.

This competition implies that robustness is not merely a matter of choosing stronger collisions. Increasing the collision angles can indeed generate correlations more rapidly per round; however, it can also amplify coherences that are most susceptible to dephasing, and it can drive the stroboscopic dynamics away from the interference condition required to build the symmetric Dicke component. In other words, overly strong collisions may overmix the excitation manifold and overshoot the target, so that noise erases the resulting coherences before they are constructively accumulated. Robust preparation therefore requires identifying collision strengths that optimally balance entanglement production against noise accumulation over the full sequence.

By rerunning the bound-constrained optimization for each decoherence strength, our procedure identifies precisely these operating points and thereby sustains high fidelities, typically at the expense of requiring additional collision rounds.

The combined results are reported in Fig.~\ref{fig:fig_robustness_analysis}. Across a broad range of shuttle failure probabilities and decoherence strengths, we find that the protocol remains effective in preparing Dicke states with the desired excitation number, with fidelities remaining high when sufficient rounds are allowed. These findings support the central practical message of this work: once collision strengths are treated as design parameters and optimized under the \emph{actual} imperfection model, collision-based Dicke-state generation can remain both deterministic and resilient, even for relatively large registers and multi-excitation targets.

\section{Discussions}\label{sec:Discussions}

\subsection{Grid Resolution and Computational Cost}
Our numerical optimization strategy combines a dense multi-start initialization on a two-dimensional grid in $(\gamma_{\mathrm{in}},\gamma_{\mathrm{sh}})$ with local refinement using L-BFGS-B. The grid spacing therefore plays a dual role: it controls the diversity of initial conditions (hence the probability of landing in a basin that contains near-global optima), and it sets the overall computational cost.

Using the baseline grid spacing of $0.2$ radians (Fig.~\ref{fig:fig_fidelity_10_qubits}), we already obtain high fidelities across Dicke states up to $n=14$ and across various accessible excitation sectors. When the grid spacing is reduced to $0.05$ radians (Fig.~\ref{fig:fig_fidelity_10_qubits_optimized}), we observe a systematic but modest improvement, which becomes more visible for larger $n$ and for intermediate excitation numbers where the optimization landscape is typically more structured. This behavior is consistent with the intuition that larger Dicke states require increasingly fine-tuned interference among many computational-basis amplitudes; consequently, small deviations from optimal collision strengths can accumulate over many rounds and become more apparent as $n$ grows.

The improvement, however, comes at a significant cost. Since the multi-start stage samples a two-dimensional domain, the number of grid points scales approximately as $\mathcal{O}(\Delta^{-2})$ with the grid spacing $\Delta$. Moreover, L-BFGS-B is executed independently from each selected grid point (and each run requires repeated forward simulations to evaluate the loss). Hence, reducing $\Delta$ enhances performance but increases the runtime approximately quadratically, in addition to any overhead from the local optimization itself. Practically, this trade-off suggests two natural extensions. First, one may use an adaptive strategy: run a coarse grid to identify promising regions, then locally refine the grid only near the best candidates. Second, one may supplement grid initialization with randomized multi-start seeding concentrated around the current incumbent best point, which often yields comparable fidelity gains at a lower computational cost than globally refining the entire grid.

\subsection{Collision Rounds Under Imperfections: Non-Monotonic Scaling}
In the robustness analysis (Fig.~\ref{fig:fig_robustness_analysis}), the optimized protocol preserves high fidelities under both missing-collision events and various decoherences acting throughout the collision sequence. An important practical question is how the required number of collision rounds scales with the imperfection strength, especially the shuttle failure probability.

While it is expected that stronger imperfections generally require more rounds than the ideal case, our numerical results show that the required round count does not necessarily increase monotonically as the imperfection parameter is swept: the observed dependence can increase, decrease, then increase again. There are at least three plausible and mutually compatible reasons for this behavior.

First, the round count is effectively a discrete resource, and the ``minimal rounds to exceed a fidelity threshold'' is therefore a step-like function of parameters even when the underlying fidelity landscape is smooth. Small changes in $(\gamma_{\mathrm{in}},\gamma_{\mathrm{sh}})$ or in the noise strength can shift the protocol from one plateau to another, producing non-monotonicities that are artifacts of discretization rather than contradictions of physical intuition.

\begin{figure}[t!]
	\includegraphics[width=1\columnwidth]{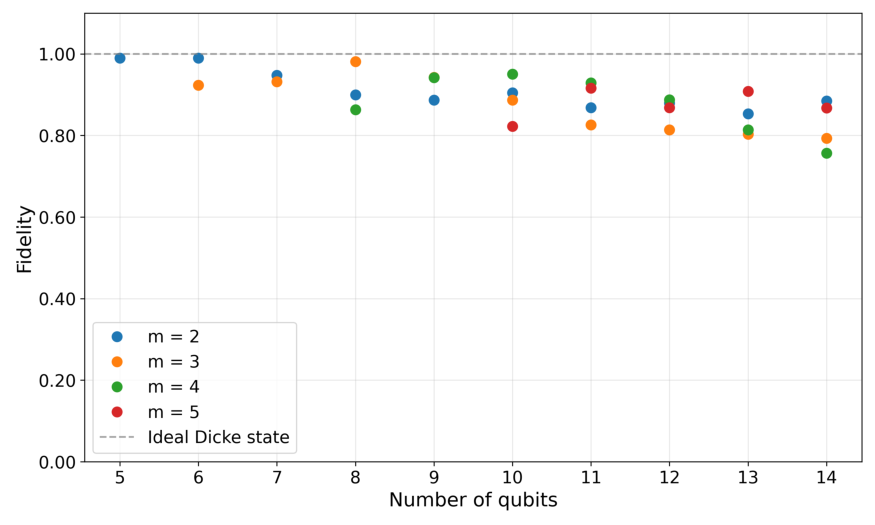}
	\caption{Fidelity between the generated and the ideal Dicke states for sizes up to 14 qubits, across various excitation sectors. Here, the simulation setup is based on a uniform grid with spacing $0.05$ radians over $\gamma_{\mathrm{in}}\in[0,\pi]$ and $\gamma_{\mathrm{sh}}\in[0.01,\pi]$.
		\label{fig:fig_fidelity_10_qubits_optimized}}
\end{figure}

Second, the optimization landscape in the presence of noise can contain multiple competing basins with comparable final fidelities but different convergence speeds. In such a situation, a slightly higher noise rate can shift the location of the best basin toward collision strengths that entangle more aggressively per round (at the expense of stronger sensitivity), which can reduce the number of rounds required to reach a target fidelity in that particular regime.

Third, the coarse multi-start grid can miss narrow basins that would provide both high fidelity and improved round efficiency at intermediate imperfection strengths. When this happens, the algorithm may settle for a near-optimal fidelity point that is not round-efficient, and the apparent round scaling becomes irregular. The improvement observed under finer grid spacing in Fig.~\ref{fig:fig_fidelity_10_qubits_optimized} supports the idea that resolution matters increasingly as the optimization problem becomes harder.

Overall a commensurate framing could be therefore: (i) the overall trend is that imperfections increase the required rounds relative to the ideal case, but (ii) the dependence can be non-monotonic at finite grid resolution because the problem is a combination of discrete stopping criteria, multi-basin optimization, and finite-resolution initialization. 

\subsection{Possible Physical Implementation}
The collision operations in our protocol are two-qubit partial-SWAP unitaries and are naturally compatible with platforms that support tunable exchange-type interactions. One particularly promising setting is superconducting circuit QED with transmon qubits, where gate times can be short compared to coherence times and where couplings can be switched on and off to emulate sequential collisions.

In this context, the role of a collision is played by activating an effective exchange interaction for a controlled duration $\tau$, yielding a partial SWAP with angle $\gamma \propto J\tau$ (with $J$ the exchange constant). State-of-the-art transmon devices exhibit coherence lifetimes on the order of microseconds, while two-qubit gate times can be on the order of tens of nanoseconds. Since one collision round in the present architecture comprises a fixed number of two-qubit activations (shuttle--register and intra-register steps), the total physical time required for tens of collisions can remain well below typical coherence times, leaving a substantial window to realize the full repeated-interaction sequence.

	Beyond analog exchange-based implementations, collision-model dynamics can also be realized in a fully digital manner on quantum computers by compiling each collision step into elementary gate sequences. Recent work has developed quantum algorithms to simulate collision models, and more broadly, Lindbladian open-system dynamics via memoryless collisions, using Hamiltonian-simulation primitives on early fault-tolerant devices, with explicit resource analyses~\cite{garg2025simulating}. Collisional representations have likewise been explored as a practical digital framework for nonequilibrium open-system phenomena such as boundary-driven transport, where the achievable accuracy is directly tied to the number of implemented collisions and the circuit depth~\cite{erbanni2023simulating}. Importantly for the present context, multipartite-collision-model algorithms have already been executed on existing noisy quantum hardware to simulate dissipative collective effects, including superradiance and subradiance, thereby providing an experimental proof-of-principle for collision-based simulations of collective dynamics~\cite{cattaneo2023quantum}.

A key experimental consideration is that the dominant noise processes act continuously throughout the collision sequence, rather than being applied only after the interactions. This is precisely the regime assessed in our robustness analysis: the entanglement generated by the repeated collisions competes with the accumulation of decoherence and other imperfections, and the optimization procedure selects collision strengths that balance these effects. From an implementation standpoint, this is advantageous because it reduces the need for postselection and projective measurements during state preparation, and it allows the protocol to be tuned to the measured noise parameters of a given device. More broadly, the same logic applies to other exchange-coupled qubit technologies where sequential activation of pairwise interactions is feasible, making collision-based Dicke-state preparation a physically transparent and flexible route toward scalable symmetric entanglement.

\section{Conclusions}\label{sec:Conclusions}
We introduced a deterministic protocol for preparing multipartite Dicke states within a collision (repeated-interaction) architecture. The scheme employs two disjoint registers coupled through excitation-preserving partial-SWAP collisions and mediated by $m$ shuttle qubits that redistribute excitations while generating multipartite correlations. In contrast to earlier collisional entanglement-generation approaches that relied on a fixed shuttle coupling and postselected projective measurements, thereby limiting the accessible target family and leading to probabilistic success, we treated both intra-register and shuttle--register collision strengths as free parameters and determined them by bound-constrained optimization that directly maximizes the target-state fidelity. This design choice enables scalable Dicke-state preparation without removing ancillary qubits by measurement and without sacrificing determinism.

We further assessed robustness under experimentally motivated imperfections that act throughout the preparation process, focusing on missing shuttle collisions and decoherence channels. In both cases, the optimization identifies collision strengths that maintain high fidelities by balancing entanglement generation against noise accumulation, typically requiring additional collision rounds relative to the ideal case. We also discussed the role of grid resolution in multi-start initialization: refining the grid improves performance, especially for larger Dicke states, but increases the computational burden approximately quadratically due to the two-parameter search.

These results position collision-based architectures as an effective and physically transparent route to deterministic symmetric multipartite entanglement. Natural future directions include adaptive initialization and refinement strategies to reduce computational overhead, optimization objectives that explicitly penalize excessive collision rounds, and experimental implementations in exchange-coupled qubit platforms where sequential pairwise interactions can be programmed with high precision.

The simulation software we developed for the optimization can be found in Ref.~\cite{nguyen2026collision}.\\

\section*{Acknowledgments}

F.O. acknowledges Tokyo International University Personal Research Fund.

\newpage
\onecolumngrid
\onecolumngrid
\section*{Appendix}

\begin{longtable}{|c|c|c|c|l|}
	\hline
	\textbf{State} & \textbf{$\gamma_{\text{sh}}$} & \textbf{$\gamma_{\text{in}}$} & \textbf{Rounds} & \textbf{$R_z$ angles (radians)} \\
	\hline
	$|D_{5}^{2}\rangle$ & 1.624 & 0.000 & 174 & \small $[-0.856, 1.413, -0.884, -0.936, 1.263]$ \\
	\hline
	$|D_{6}^{2}\rangle$ & 3.037 & 0.263 & 106 & \small $[-1.620, -1.847, 0.959, 0.876, 0.836, 0.795]$ \\
	\hline
	$|D_{6}^{3}\rangle$ & 0.475 & 0.243 & 3 & \small [1.749, 1.044, 0.442, -1.775, -0.729, -0.731] \\
	\hline
	$|D_{7}^{2}\rangle$ & 0.010 & 1.678 & 126 & \small [-2.034, -2.016, 0.814, 0.784, 0.734, 0.829, 0.889] \\
	\hline
	$|D_{7}^{3}\rangle$ & 3.008 & 0.400 & 95 & \small [1.603, 1.260, 1.586, -1.094, -1.147, -1.107, -1.103] \\
	\hline
	$|D_{8}^{2}\rangle$ & 0.431 & 0.000 & 2 & \small [2.323, 1.414, -1.815, -0.815, 0.067, -1.293, -0.365, 0.483] \\
	\hline
	$|D_{8}^{3}\rangle$ & 0.041 & 1.300 & 179 & \small [1.638, 1.665, 1.534, -1.017, -0.943, -1.001, -0.945, -0.931] \\
	\hline
	$|D_{8}^{4}\rangle$ & 2.736 & 3.142 & 4 & \small [-1.992, -1.362, -0.793, -0.321, 1.992, 0.793, 1.362, 0.321] \\
	\hline
	$|D_{9}^{2}\rangle$ & 0.406 & 0.000 & 2 & \small [2.488, 1.572, -1.910, -0.975, -0.132, -1.425, -0.546, 0.269, 0.659] \\
	\hline
	$|D_{9}^{3}\rangle$ & 0.011 & 1.614 & 177 & \small [-1.845, -1.824, -1.803, 0.773, 0.913, 1.034, 0.783, 0.923, 1.045] \\
	\hline
	$|D_{9}^{4}\rangle$ & 0.061 & 2.685 & 192 & \small [1.254, 1.283, 1.355, 1.420, -1.134, -1.118, -1.019, -1.023, -1.019] \\
	\hline
	$|D_{10}^{2}\rangle$ & 3.067 & 1.279 & 99 & \small [2.815, -3.345, 1.273, 0.528, -0.409, -0.947, 1.097, 0.398, -0.459, -0.951] \\
	\hline
	$|D_{10}^{3}\rangle$ & 0.010 & 1.760 & 153 & \small [-2.041, -2.020, -2.000, 0.818, 0.855, 0.895, 0.865, 0.875, 0.872, 0.881] \\
	\hline
	$|D_{10}^{4}\rangle$ & 3.070 & 1.953 & 144 & \small [-1.478, -1.476, -1.582, -1.755, 1.132, 1.056, 0.992, 1.123, 1.016, 0.973] \\
	\hline
	$|D_{10}^{5}\rangle$ & 2.797 & 3.131 & 5 & \small [-2.140, -1.617, -1.105, -0.648, -0.267, 2.147, 1.123, 1.600, 0.637, 0.269] \\
	\hline
	$|D_{11}^{2}\rangle$ & 0.367 & 0.000 & 3 & \small [-2.936, 2.423, -1.516, -0.683, 0.088, 0.826, -1.089, -0.292, 0.461, 1.184, 1.535] \\
	\hline
	$|D_{11}^{3}\rangle$ & 2.801 & 3.142 & 4 & \small [-2.549, -1.851, -1.142, 2.078, 1.226, 0.457, -0.238, 1.641, 0.833, 0.100, -0.555] \\
	\hline
	$|D_{11}^{4}\rangle$ & 0.099 & 0.217 & 121 & \small [1.530, 1.664, 1.751, 1.822, -1.055, -1.042, -1.042, -0.844, -0.864, -0.937, -0.983] \\
	\hline
	$|D_{11}^{5}\rangle$ & 3.045 & 2.724 & 170 & \small [1.406, 1.317, 1.270, 1.292, 1.388, -1.115, -1.128, -1.133, -1.084, -1.097, -1.115] \\
	\hline
	$|D_{12}^{2}\rangle$ & 2.786 & 0.112 & 3 & \small [2.745, -2.264, 1.645, 0.905, 0.124, -0.575, -1.459, 1.244, 0.529, -0.225, -0.904, -1.765] \\
	\hline
	$|D_{12}^{3}\rangle$ & 0.324 & 0.000 & 4 & \small [2.664, 1.978, 1.249, -2.138, -1.339, -0.606, 0.071, -1.729, -0.965, -0.260, 0.388, 0.687] \\
	\hline
	$|D_{12}^{4}\rangle$ & 0.017 & 1.637 & 117 & \small [-1.908, -1.879, -1.851, -1.823, 0.680, 0.868, 1.020, 1.131, 0.696, 0.884, 1.036, 1.147] \\
	\hline
	$|D_{12}^{5}\rangle$ & 3.039 & 2.691 & 180 & \small [-1.877, -2.075, -2.194, -2.358, -1.732, 1.470, 1.450, 1.452, 1.445, 1.450, 1.488, 1.482] \\
	\hline
	$|D_{13}^{2}\rangle$ & 0.338 & 0.000 & 3 & \small [-2.777, 2.579, -1.768, -1.012, -0.299, 0.388, 1.058, -1.382, -0.652, 0.047, 0.725, 1.386, 1.708] \\
	\hline
	$|D_{13}^{3}\rangle$ & 2.832 & 3.142 & 4 & \small [-2.770, -2.096, -1.350, 2.196, 1.442, 0.742, 0.086, -0.524, 1.812, 1.086, 0.408, -0.225, -0.808] \\
	\hline
	$|D_{13}^{4}\rangle$ & 0.010 & 1.648 & 167 & \small [-2.017, -2.000, -2.045, -2.028, 0.745, 0.839, 0.961, 1.019, 0.888, 0.893, 0.906, 0.919, 0.919] \\
	\hline
	$|D_{13}^{5}\rangle$ & 0.015 & 1.643 & 193 & \small [2.131, 2.156, 2.122, 2.150, 2.179, -1.452, -1.384, -1.299, -1.251, -1.444, -1.376, -1.289, -1.241] \\
	\hline
	$|D_{14}^{2}\rangle$ & 0.327 & 3.024 & 3 & \small [-2.610, 2.416, -1.855, -1.192, -0.467, 0.199, 0.812, 1.635, -1.493, -0.846, -0.141, 0.511, 1.112, 1.919] \\
	\hline
	$|D_{14}^{3}\rangle$ & 2.845 & 3.142 & 4 & \small [2.964, -2.656, -1.893, 1.805, 1.090, 0.419, -0.215, -0.814, 1.441, 0.749, 0.097, -0.519, -1.098, -1.370] \\
	\hline
	$|D_{14}^{4}\rangle$ & 2.860 & 3.142 & 5 & \small [-2.694, -2.156, -1.519, -0.993, 2.245, 1.522, 0.840, 0.212, -0.350, 1.879, 1.175, 0.519, -0.079, -0.600] \\
	\hline
	$|D_{14}^{5}\rangle$ & 0.015 & 0.066 & 167 & \small [3.055, -9.562, 3.037, 3.070, 3.103, -0.315, -0.311, -0.308, -0.304, -0.291, -0.290, -0.293, -0.297, -0.295] \\
	\hline
	\caption{Optimization results for deterministic preparation of Dicke states.}
	\label{tab:optimization_results_d11_14}
\end{longtable}

\twocolumngrid
\twocolumngrid

%%%%%%%%%%%%%%%%%%%%%%%%%%%%%%%%%%%%%%%%%%%%%%%%%%%%%%%%%%%%%%%%%%
%%%%%%%%%%%%%%%%%%%%%%%%%%%%%%%%%%%%%%%%%%%%%%%%%%%%%%%%%%%%%%%%%%%
\end{document}